\documentclass[12pt]{iopart}
\usepackage{iopams}
\usepackage{setstack}
\usepackage{psfig}

\newcommand{\cP}{\ensuremath{\mathcal{P}}}
\newcommand{\cT}{\ensuremath{\mathcal{T}}}
\newcommand{\half}{\mbox{$\textstyle{\frac{1}{2}}$}}

\begin{document}
\rightline{preprint LA-UR-06-6814}
\title[Complex Trajectories of a Simple Pendulum]{Complex Trajectories of a
Simple Pendulum}

\author[Bender, Holm, and Hook]{Carl~M~Bender$^*$\footnote{Permanent address:
Department of Physics, Washington University, St. Louis MO 63130, USA},
Darryl~D~Holm$^\dag$, and Daniel~W~Hook$^{\ddag}$}

\address{${}^*$Center for Nonlinear Studies, Los Alamos National Laboratory,
Los Alamos, NM 87545, USA\\ {\footnotesize email: cmb@wustl.edu}}
\vspace{.2cm}
\address{${}^{\dag}$Department of Mathematics, Imperial College, London SW7 2AZ,
UK\\ {\footnotesize email: d.holm@ic.ac.uk}\\and\\
Computer and Computational Science, Los Alamos National Laboratory,\\
MS D413 Los Alamos, NM 87545, USA \\ {\footnotesize email: dholm@lanl.gov}}
\vspace{.2cm}
\address{${}^{\ddag}$Blackett Laboratory, Imperial College, London SW7 2BZ, UK
\\{\footnotesize email: d.hook@imperial.ac.uk}}

\begin{abstract}
The motion of a classical pendulum in a gravitational field of strength $g$ is
explored. The complex trajectories as well as the real ones are determined. If
$g$ is taken to be imaginary, the Hamiltonian that describes the pendulum
becomes $\cP\cT$-symmetric. The classical motion for this $\cP\cT$-symmetric
Hamiltonian is examined in detail. The complex motion of this pendulum in
the presence of an external periodic forcing term is also studied.
\end{abstract}
\submitto{\JPA}
\section{Introduction}
\label{s1}
Many papers have been written on $\cP\cT$-symmetric quantum-mechanical
Hamiltonians. In nearly all cases these Hamiltonians have had rising potentials
and energies that are all discrete and {\it real}. An example of such a class of
Hamiltonians is \cite{r1,r2,r3} 
\begin{equation}
H=\half p^2+x^2(ix)^\epsilon\qquad(\epsilon\geq0).
\label{e1}
\end{equation}
In a few instances quantum-mechanical Hamiltonians having real continuous
spectra have been examined \cite{r4}.

For the case of $\cP\cT$-symmetric quantum-mechanical Hamiltonians having
discrete spectra, the associated classical-mechanical systems have also been
studied \cite{r2,r5,r6,r7}. It has been found that the classical trajectories
associated with real energies are $\cP\cT$-symmetric (they are symmetric under
reflection about the imaginary axis) and exhibit remarkably interesting
topological properties. These trajectories can visit multiple sheets of a
Riemann surface. However, the classical mechanics associated with quantum
systems having continuous spectra have not yet been examined. Thus, in this
paper we explore the classical trajectories of the periodic potential analyzed
in Ref.~\cite{r4}.

\begin{figure}[th]\vspace{2.8in}
\includegraphics{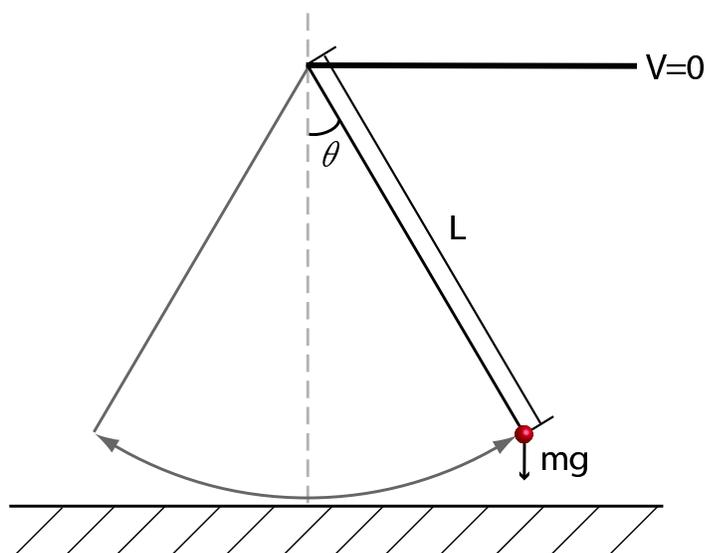}
\caption{Configuration of a simple pendulum of mass $m$ in a uniform
gravitational field of strength $g$. The length of the string is $L$. The
pendulum swings through an angle $\theta$. We define the potential energy to be
$0$ at the height of the pivot.
\label{f1}}
\end{figure}

Consider a classical-mechanical system consisting of a simple pendulum with a
bob of mass $m$ and a string of length $L$ in a uniform gravitational field of
magnitude $g$ (see Fig.~\ref{f1}). The gravitational potential energy of the
system is defined to be zero at the height of the pivot point of the string. The
pendulum bob swings through an angle $\theta$. Therefore, the horizontal and
vertical cartesian coordinates $X$ and $Y$ are given by
\begin{equation}
X=L\sin\theta,\qquad Y=-L\cos\theta,
\label{e2}
\end{equation}
which gives velocities
\begin{equation}
\dot{X}=L\dot{\theta}\sin\theta,\qquad\dot{Y}=-L\dot{\theta}\cos\theta.
\label{e3}
\end{equation}
We define the potential and kinetic energies $V$ and $T$ as
\begin{equation}
V=-mgL\cos\theta,\qquad T=\half m(\dot{X}^2+\dot{Y}^2)=\half mL^2\dot{\theta}^2.
\label{e4}
\end{equation}
The Hamiltonian $H=T+V$ for the pendulum is therefore
\begin{equation}
H=\half mL^2\dot{\theta}^2-mgL\cos\theta.
\label{e5}
\end{equation}
Without loss of generality we set $m=1$ and $L=1$, and for consistency with the
notation used in earlier papers on $\cP\cT$-symmetric Hamiltonians we relabel
$\theta\rightarrow x$ to get
\begin{equation}
H=\half p^2-g\cos x,
\label{e6}
\end{equation}
where $p=\dot{x}$. The classical equations of motion for this Hamiltonian read
\begin{equation}
\dot{x}=\frac{\partial H}{\partial p}=p,\qquad\dot{p}=-\frac{\partial H}{
\partial x}=-g\sin x.
\label{e7}
\end{equation}
The Hamiltonian $H$ for this system is a constant of the motion and thus the
energy $E$ is a time-independent quantity. For most of this paper we take $E$ to
be real because the energy levels of the corresponding quantum-mechanical
system are real.

This paper is organized as follows: In Sec.~\ref{s2} we study the complex
trajectories for $H$ in (\ref{e6}) when $E$ and $g$ are real. In Sec.~\ref{s3}
we study the complex trajectories for $H$ when $E$ is real but $g$ is imaginary.
In Sec.~\ref{s4} we examine briefly the complex chaotic motion that results when
the energy $E$ of the pendulum is allowed to be complex and also when the
pendulum is subject to a periodic external driving force.

\section{Pendulum in a Real Gravitational Field}
\label{s2}

In this section we examine the solutions to Hamilton's equations (\ref{e7})
for the case in which the gravitational field strength $g$ is real. Without loss
of generality we take $g=1$. There are three cases to consider: (1) the complex
closed trajectories that result when $|E|<1$, (2) the complex open trajectories
that occur when $E>1$, and (3) the complex closed trajectories that we obtain
when $E<-1$.

\subsection{Closed periodic trajectories: $|E|<1$.}

\begin{figure}[th]\vspace{2.55in}
\includegraphics{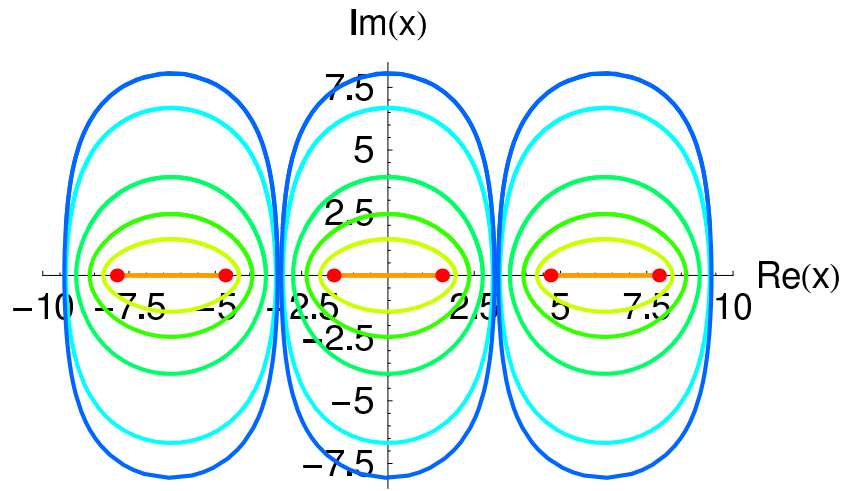}
\caption{Complex classical trajectories for the simple pendulum in (\ref{e6}).
These trajectories satisfy the equations of motion in (\ref{e7}) with $g=1$
and $E=0$. Note that all trajectories are closed periodic orbits and lie in
cells of width $2\pi$ that entirely fill the complex plane. The usual physically
observed motion of a swinging pendulum is represented by the line segments
joining pairs of turning points on the real axis.
\label{f2}}
\end{figure}

When the energy $E$ of a simple pendulum lies in the range $-1\leq E\leq1$, the
physically observed motion is that of a swinging pendulum. For the example
considered in this subsection we choose $E=0$. Since a {\it classical turning
point} $x_0$ is the solution to the equation $V(x_0)=E$, the classical turning
points for this case satisfy $\cos x_0=0$. Thus,
\begin{equation}
x_0=\pi/2+n\pi \quad(n\in{\mathbb Z}).
\label{e8}
\end{equation}
The observed classical trajectory for this pendulum is represented by $x(t)$ on
the real-$x$ axis. This trajectory oscillates on the real axis between the
turning points at $x_0=\pm\half\pi$. However, the full array of complex
solutions to the classical equations of motion is much richer (see
Fig.~\ref{f2}). In this figure we see that all classical trajectories are
closed and periodic and lie in cells of width $2\pi$ that fill the entire
complex-$x$ plane. As a consequence of Cauchy's theorem, all of the closed
orbits have exactly the same period; namely, the period of a real physical
pendulum that undergoes periodic motion on the real axis. This is because the
period $T$ for a particle of energy $E$ for a Hamiltonian of the form $H=\half
p^2+V(x)$ is given in terms of a closed complex contour integral: $T=\oint_C dx/
\sqrt{2[E-V(x)]}$, where $C$ is the complex closed path of the particle. Since
the only singularity that $C$ encloses is the square-root branch cut joining the
turning points, the value of $T$ is independent of $C$.

\begin{figure}[th]\vspace{2.6in}
\includegraphics{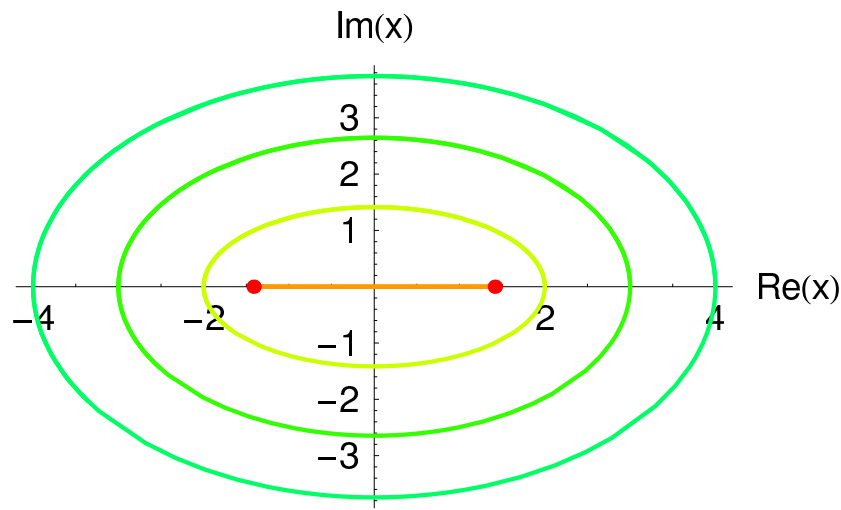}
\caption{Complex trajectories of the simple harmonic oscillator, whose
Hamiltonian is $H=\half p^2+\half x^2$ with $E=1$. These orbits are exact 
ellipses. These ellipses closely approximate the complex orbits in the vicinity
of the turning points in Fig.~\ref{f2}.
\label{f3}}
\end{figure}

Note that the orbits in Fig.~\ref{f2} that lie near the turning points are
approximately elliptical and resemble the exactly elliptic complex trajectories
of the simple harmonic oscillator, whose Hamiltonian is $H=\half p^2+\half x^2$.
These orbits are displayed in Fig.~\ref{f3}.

\subsection{Open trajectories: $E>1$.}

When the energy $E$ of a simple pendulum is sufficiently large (here, $E>1$),
the physically observed motion is that of a rotating pendulum. The classical
turning points for this case are no longer on the real axis because there are no
real solutions to $\cos x_0=E$ with $E>1$. The rotating pendulum has no real
turning points because its real trajectory never turns back. To find the complex
turning points, we replace $x_0$ by $a+ib$ and get two equations:
\begin{equation}
\sin a\sinh b=0,\qquad\cos a\cosh b=-E.
\label{e9}
\end{equation}
For the example considered in this subsection we choose $E=\cosh1$, and thus
the turning points are situated at $a=(2k+1)\pi$, where $k\in{\mathbb Z}$, and
$b=\pm1$.

\begin{figure}[th]\vspace{2.6in}
\includegraphics{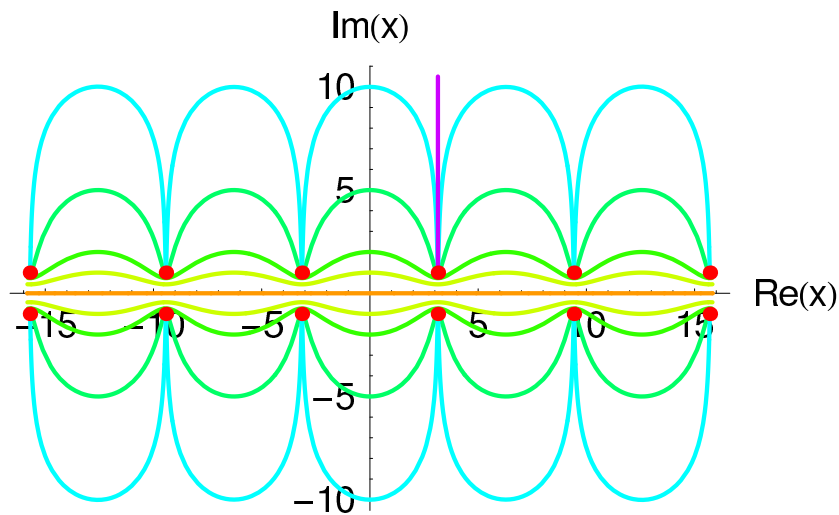}
\caption{Trajectories in the complex-$x$ plane for the rotating pendulum. For
this case the pendulum has energy $E=\cosh1$ and the gravitational field
strength has the value $g=1$. Note that the trajectory on the real axis
corresponds to a physical pendulum undergoing continuous circular motion.
All of the complex trajectories are open. The trajectories that terminate at
turning points run vertically off to complex infinity. One such trajectory that
starts at $x_0=\pi+i$ is shown.
\label{f4}}
\end{figure}

The complex trajectories for this case are shown in Fig.~\ref{f4}. Note that the
trajectories fill the entire complex plane and that there are no closed orbits.
The turning points have a strong influence on the path of the complex pendulum
bob. The trajectories in the upper-half complex-$x$ plane, for example, go just
below and veer around the turning points above the real axis. The turning points
seem to attract the trajectories in much the same way as was found in
Ref.~\cite{r6}.

\begin{figure}[th]\vspace{3.85in}
\includegraphics{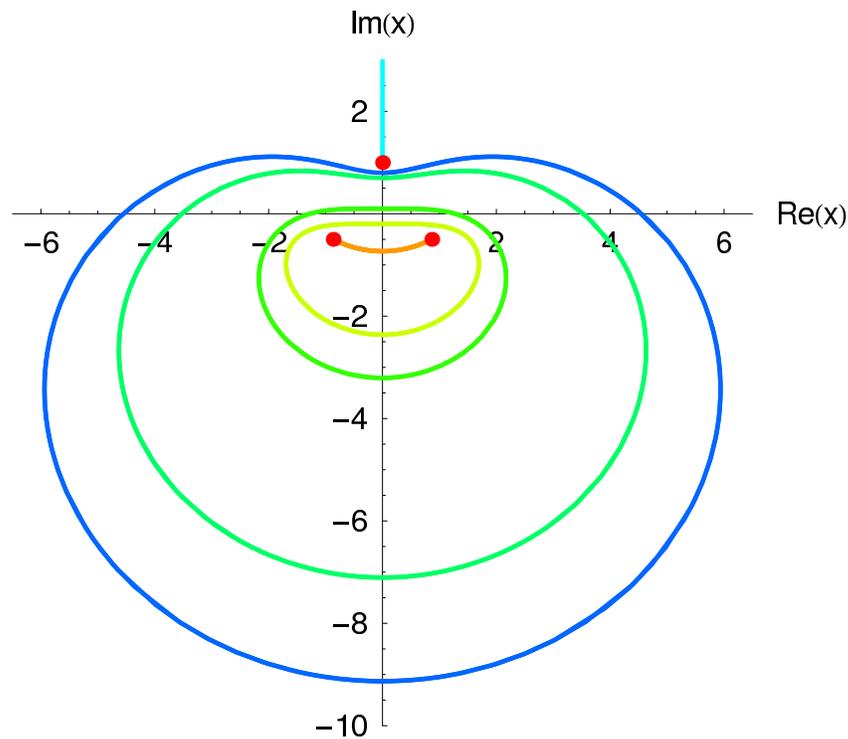}
\caption{Complex trajectories for the Hamiltonian $H=\half p^2+ix^3$ for $E=1$.
All trajectories are closed periodic orbits except for the trajectory beginning
at $x_0=i$. A particle released from this turning point runs off to infinity
in a finite time.
\label{f5}}
\end{figure}

Note that the trajectories that terminate at turning points run vertically off
to complex infinity. This feature is also a property of the Hamiltonian $H=\half
p^2+ix^3$, which was studied in detail in Ref.~\cite{r2}. The complex
trajectories for this Hamiltonian are shown in Fig.~\ref{f5}. There are three
turning points, and the trajectory beginning at the turning point on the
imaginary axis goes up the imaginary axis to infinity.

One may ask how long it takes for a particle beginning at a turning point, say,
the turning point at $\pi+i$, to reach infinity. The time $T$ for this motion is
represented by the following integral:
\begin{equation}
T=\frac{1}{\sqrt{2}}\int_{x=i+\pi}^{i\infty+\pi}\frac{dx}{\sqrt{E+\cos x}}.
\label{e10}
\end{equation}
A numerical evaluation of this integral gives $T=1.97536\cdots$. Thus, the
particle reaches the point at complex infinity in a finite time.

\subsection{Closed trajectories: $E<-1$}

\begin{figure}[th]\vspace{2.65in}
\includegraphics{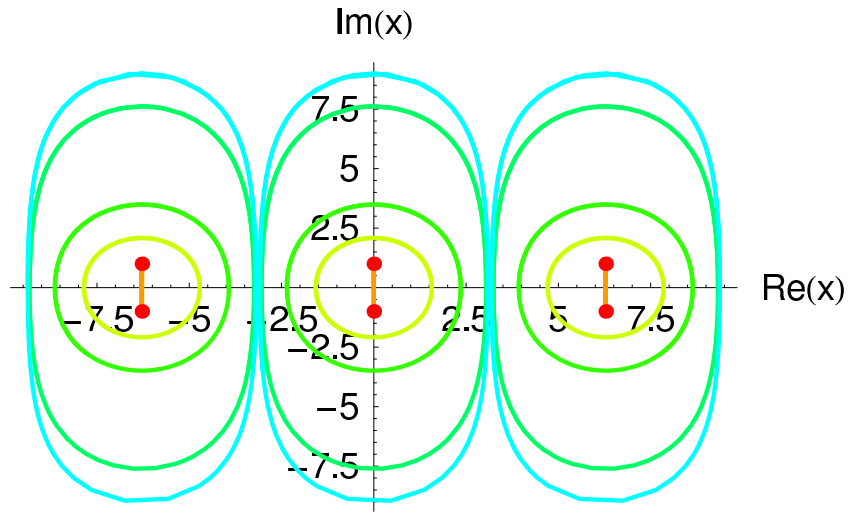}
\caption{Complex trajectories for the pendulum in (\ref{e6}) with $g=1$ and
$E=-\cosh1$. The turning points come in complex-conjugate pairs and the
trajectories are all closed orbits.
\label{f6}}
\end{figure}

There are no physically observable trajectories for this choice of the energy.
However, the complex trajectories in this case can be calculated numerically
(see Fig.~\ref{f6}) and we find that they are all closed orbits that lie in
periodic cells of width $2\pi$. The turning points come in complex-conjugate
pairs. In Fig.~\ref{f6} we have chosen $E=-\cosh1$ and the turning points are
situated at $x_0=a+ib$, where $a=2k\pi$ ($k\in{\mathbb Z}$) and $b=\pm1$.

\section{Pendulum in an Imaginary Gravitational Field}
\label{s3}

If we take the gravitational field strength $g$ in (\ref{e6}) to be imaginary,
the Hamiltonian becomes $\cP\cT$ symmetric, where $\cP$ reflection consists
of $x\rightarrow \half n\pi-x$. Setting $g=i$, the Hamiltonian becomes
\begin{equation}
H=p^2/2-{\rm i}\cos x
\label{e11}
\end{equation}
and the turning points are located at $x_0=a+ib$, where
\begin{equation}
\cos a\cosh b=0,\qquad-\sin a\sinh b=E.
\label{e12}
\end{equation}

\begin{figure}[th]\vspace{2.5in}
\includegraphics{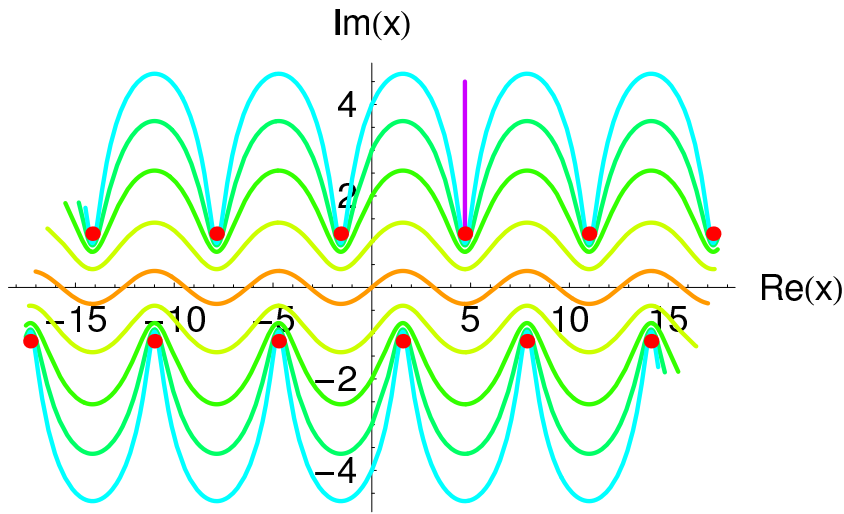}
\caption{Complex trajectories for the pendulum in an imaginary gravitational
field of strength $g=i$ and positive real energy $E=\sinh 1$.
\label{f7}}
\end{figure}

\begin{figure}[th]\vspace{2.65in}
\includegraphics{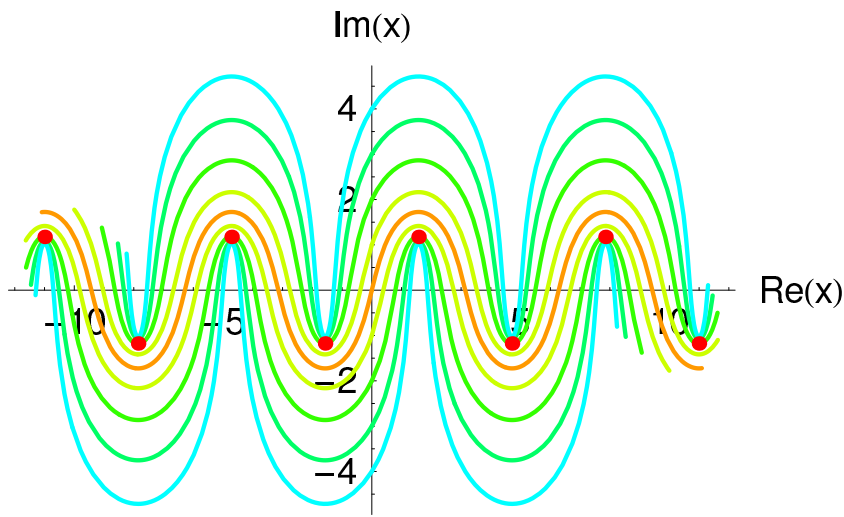}
\caption{Complex trajectories for the pendulum in an imaginary gravitational
field of strength $g=i$ and negative real energy $E=-\sinh 1$. Trajectories that
begin at the turning points run off to complex infinity in the positive and
negative imaginary directions as in Figs.~\ref{f4}, \ref{f5}, and \ref{f7}.
\label{f8}}
\end{figure}

There are now two cases to consider, $E>0$ and $E<0$. In Fig.~\ref{f7} we take
$E=\sinh1$ and find turning points at $x_0=\left(n+\half\right)\pi+i(-1)^n$ ($n
\in{\mathbb Z}$) and in Fig.~\ref{f8} we take $E=-\sinh1$ and find turning
points at $x_0=\left(n+\half\right)\pi+i(-1)^{n+1}$ ($n \in{\mathbb Z}$). Like
the complex trajectories in Fig.~\ref{f4}, the trajectories in these two
figures are not closed and trajectories beginning at turning points run
vertically off to infinity. In Fig.~\ref{f7} one such trajectory is shown.
This trajectory begins at $\frac{3}{2}\pi+i$. The time $T$ for a particle to
travel from this turning point to infinity is given by
\begin{equation}
T=\frac{1}{\sqrt{2}}\int_{x=i+3\pi/2}^{i\infty+3\pi/2}\frac{dx}{\sqrt{E+i\cos x}
}.
\label{e13}
\end{equation}
After a simple change of variables this integral becomes
\begin{equation}
T=\frac{1}{\sqrt{2}}\int_{s=1}^\infty\frac{ds}{\sqrt{\sinh s-\sinh1}}=\frac{2}
{\sqrt{e}}K(-1/e^2)=1.84549\cdots,
\label{e14}
\end{equation}
where $K$ is the complete elliptic function. Thus, the classical particle
reaches complex infinity in finite time.

\section{Further Considerations}
\label{s4}

There are a number of interesting ways to extend and generalize the
investigations in the previous two sections.

\begin{figure}[th]\vspace{2.65in}
\includegraphics{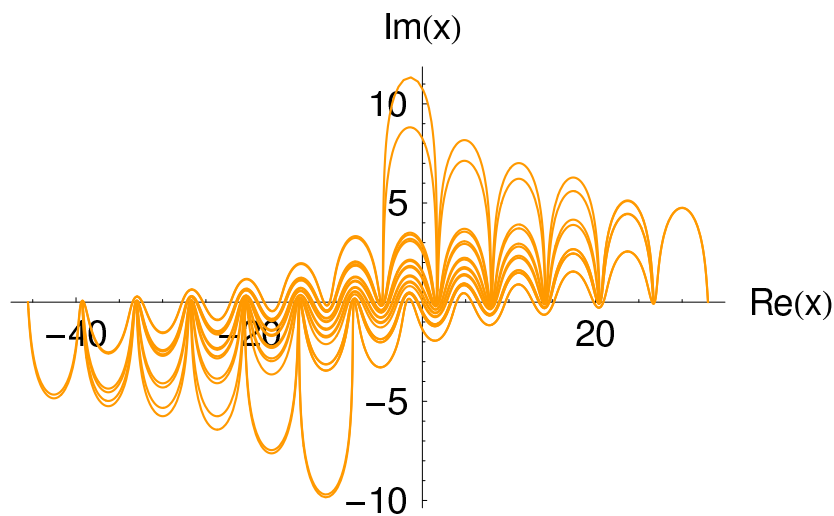}
\caption{Complex trajectories for a pendulum in an imaginary gravitational
field of strength $g=i$ and complex energy $E=i$. In this case the trajectories
are not closed and not spatially periodic.
\label{f9}}
\end{figure}

In Figs.~\ref{f7} and \ref{f8} the energy for the pendulum in an imaginary
gravitational field was taken to be real. This is because a quantum-mechanical
$\cP\cT$-symmetric Hamiltonian has the virtue of possessing real energy levels.
However, one might ask what happens if one takes the energy to be complex. In
this case the complex trajectories are not spatially or temporally periodic,
even though the turning points {\it are} periodic and are separated horizontally
by $2\pi$. The behavior of the trajectories is illustrated in Fig.~\ref{f9},
where we have taken the energy to be $E=i$.

\begin{figure}[th]\vspace{2.6in}
\includegraphics{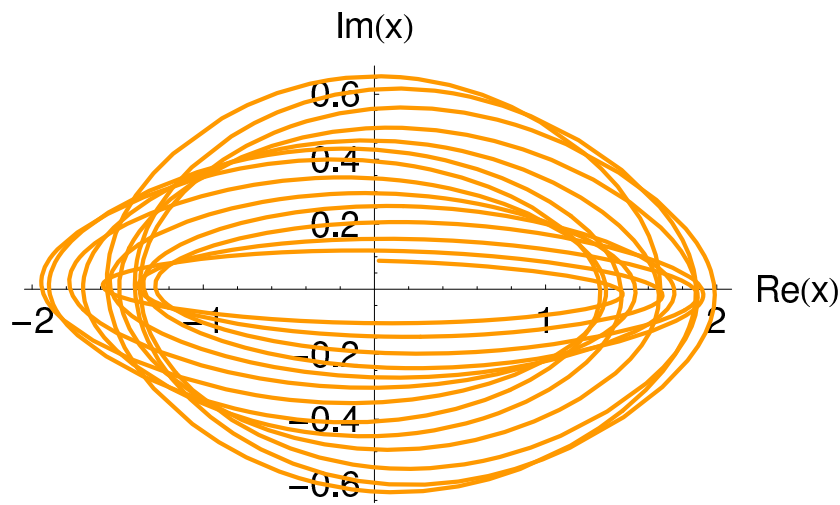}
\caption{Complex trajectories for a pendulum in a gravitational field of
strength $g=1$ and subject to an external periodic driving force. Here,
the driving force has the form $0.2\sin(0.1t)$ and the initial condition is
taken to be $x=\half\pi+0.1$. Ordinarily, such an initial condition would
lead to a complex trajectory like that in Fig.~\ref{f2}. For short time (up
to $t=100$ in this figure) the trajectory remains confined to the central
cell in Fig.~\ref{f2}.
\label{f10}}
\end{figure}

\begin{figure}[th]\vspace{2.6in}
\includegraphics{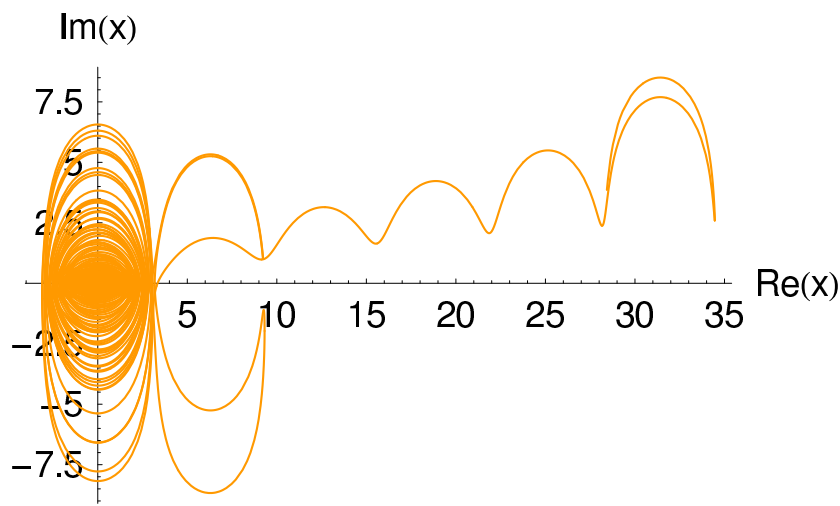}
\caption{Same as in Fig.~\ref{f10} but now the trajectory is allowed to
continue until $t=650$. Eventually, the trajectory excapes from its cell and
begins to wander into nearby cells. It gains complex energy and it begins to
resemble the trajectory in Fig.~\ref{f9}.
\label{f11}}
\end{figure}

\begin{figure}[th]\vspace{2.6in}
\includegraphics{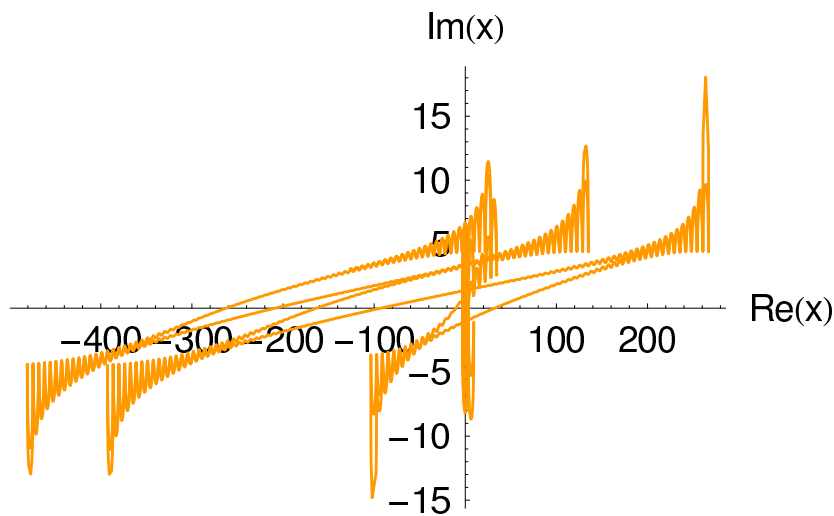}
\caption{Same as in Fig.~\ref{f11} but now the time runs up to $t=1000$.
Note that the complex trajectory undergoes extreme fluctuations.
\label{f12}}
\end{figure}

Another way to extend these investigations is to study what happens when there
is an external periodic driving force of the form $\epsilon\sin(\omega t)$
\cite{r8,r9,r10,r11}. We have taken $\epsilon=0.2$ and $\omega=0.1$ and in
Figs.~\ref{f10}, \ref{f11}, and \ref{f12} we have examined the complex
trajectories up to times $t=100$, $t=650$, and $t=1000$. The starting energy for
these figures is taken to be $0$ and the value of $g$ is $1$. The initial value
of $x$ is taken to be $\half\pi+0.1$,which would lead to a complex trajectory if
there were no driving term. We see that for short times the trajectory is noisy,
but seems to be confined to the central cell in Fig.~\ref{f2}. However, after a
sufficiently long time the trajectory wanders into adjacent cells, as can be
seen in Figs.~\ref{f11} and \ref{f12}. This result suggests it might be
productive to investigate complex trajectories of dynamical systems whose real
trajectories exhibit chaos, especially in the $\cP\cT$-symmetric case. One could
begin with the complex trajectories associated with real homoclinic chaos, as in
the present case. In addition, the rich behavior seen in the complex spatial
trajectories of $\cP \cT$-symmetric systems may be expected to be the projection
onto the two-dimensional complex plane of interesting behavior in
four-dimensional complex phase space. For example, the phase-space trajectories
of $\cP\cT$-symmetric complex systems whose real parts are chaotic might exhibit
a complex Hamiltonian analog of Arnold diffusion, or some other kind of
Hamiltonian diffusion.

\vspace{0.5cm}
\begin{footnotesize}
\noindent
We thank D.~C.~Brody for many useful discussions. CMB thanks the Theoretical
Physics Group at Imperial College, London, for its hospitality. As an Ulam
Scholar, CMB receives financial support from the Center for Nonlinear Studies at
the Los Alamos National Laboratory and he is supported in part by a grant from 
the U.S. Department of Energy. The work of DDH was supported by the Royal
Society of London and by the U.S. Department of Energy Office of Science Applied
Mathematical Research.
\end{footnotesize}

\end{document}